\documentclass[12pt,a4paper]{article}
\usepackage{graphicx}
\usepackage{amsmath}
\usepackage{amsfonts}
\usepackage{amssymb}
\usepackage{epstopdf}
\usepackage{bm}

\newcommand{\trac}[2]{{\displaystyle\frac{#1}{#2}}}

\newcommand{\mg}{m_{3/2}}
\newcommand\ee{\end{equation}}
\newcommand\be{\begin{equation}}
\newcommand\eea{\end{eqnarray}}
\newcommand\bea{\begin{eqnarray}}

\renewcommand\({\left(}
\renewcommand\){\right)}

\usepackage{epsfig,multicol}
\usepackage{latexsym}
\usepackage{color}

\usepackage{amssymb,amsmath}

\newcommand{\dis}[1]{\begin{equation}\begin{split}#1\end{split}\end{equation}}

\newcommand{\eq}[1]{Eq.~(\ref{#1})}

\newcommand\lsim{\mathrel{\rlap{\lower4pt\hbox{\hskip1pt$\sim$}}
    \raise1pt\hbox{$<$}}}
\newcommand\gsim{\mathrel{\rlap{\lower4pt\hbox{\hskip1pt$\sim$}}
    \raise1pt\hbox{$>$}}}

\newcommand{\newc}{\newcommand}

\newc{\mhalf}{m_{1/2}}      \newc{\mzero}{m_0}

\newc{\tanb}{\tan\beta}
\newc{\azero}{A_0}
\newc{\at}{A_t} \newc{\abot}{A_b} \newc{\atau}{A_\tau} 

\newc{\bmu}{B\mu}           \newc{\sgn}{{\rm sgn}}
\newc{\mone}{M_1}           \newc{\mtwo}{M_2}

\newc{\charone}{\chi_1^\pm} \newc{\mcharone}{m_{\chi_1^\pm}}

\newc{\hl}{h}               \newc{\mhl}{m_{\hl}}
\newc{\hh}{H}               \newc{\mhh}{m_{\hh}}
\newc{\ha}{A}               \newc{\mha}{m_{\ha}}
\newc{\hc}{H^{\pm}}         \newc{\mhc}{m_{\hc}}

\newc{\mw}{m_{W}}      \newc{\mz}{m_{Z}}

\newc{\mgut}{M_{\rm GUT}}

\newc{\mplanck}{M_{\rm Pl}}      
\newc{\mpl}{M_{\rm Pl}}
\newc{\msusy}{M_{\rm SUSY}}      \newc{\ms}{M_{\rm S}}

\newc{\VEV}[1]{\langle #1 \rangle}

\newc{\xf}{x_f}
\newc\vrel{v_{\rm rel}}

\newcommand\stau{{\widetilde{\tau}}}

\newcommand\mstau{m_{\stau}}

\newcommand\treh{T_{\rm RH}}

\newc{\abund}{\Omega h^2}
\newc{\abundchi}{\Omega_\chi h^2}
\newc{\abundcdm}{\Omega_{{\rm CDM}} h^2}

\newc{\omeganlsp}{\Omega_{{\rm NLSP}}}  \newc{\abundnlsp}{\Omega_{\rm NLSP}h^2}
\newc{\omegalsp}{\Omega_{{\rm LSP}}}  \newc{\abundlsp}{\Omega_{\rm LSP}h^2}
\newc{\omegawmap}{\Omega_{{\rm WMAP}}}\newc{\abundwmap}{\Omega_{\rm WMAP}h^2}
\newc{\omegagravitinio}{\Omega_{{\gravitino}}}
\newc{\abundgravitino}{\Omega_{\gravitino}h^2}
\newc{\ynlsp}{Y_{{\rm NLSP}}}            \newc{\taunlsp}{\tau_{{\rm NLSP}}}
\newc{\nnlsp}{n_{{\rm NLSP}}}            \newc{\mnlsp}{m_{{\rm NLSP}}}
\newc{\mlsp}{m_{{\rm LSP}}}
\newc{\ylsp}{Y_{{\rm LSP}}}        
\newc{\nx}{n_{X}}                        \newc{\yx}{Y_{X}}
\newc{\mx}{m_{X}}                        \newc{\taux}{\tau_{X}}

\newc{\rhocrit}{\rho_{crit}}
\newc{\rhochi}{\rho_{\chi}}

\newcommand\tev{\,\mbox{TeV}}
\newcommand\gev{\,\mbox{GeV}}
\newcommand\mev{\,\mbox{MeV}}

\newcommand\mpc{\,\mbox{Mpc}}
\newc\gbar{{\overline{g}}}

\newcommand\cm{\,\mbox{cm}}
\newcommand\km{\,\mbox{km}}
\newcommand\kpc{\,\mbox{kpc}}

\newc{\ra}{\rightarrow}

\renewcommand\({\left(}
\renewcommand\){\right)}
\renewcommand\[{\left[}

\newcommand\gravitino{\widetilde{G}}    

\newcommand{\bfrac}[2]{{\left(\frac{#1}{#2} \right)  }}

\newc{\gstar}{g_\ast}           \newc{\gsstar}{g_{s\ast}}

       \def\pslash{\not{\hbox{\kern-2.3pt $p$}}}
       \def\kslash{\not{\hbox{\kern-2.3pt $k$}}}
       \def\qslash{\not{\hbox{\kern-2.3pt $q$}}}
       \def\ddslash{\not{\hbox{\kern-2.3pt $d$}}}
       \def\prtslash{\not{\hbox{\kern-2.3pt $\partial$}}}

\newcommand\jcap[3] 
		{{\it J.\ Cosmol.\ Astrop.\ Phys.\ }{\bf #1} (#2) #3}

\newcommand{\amusm}{a_{\mu}^{\text{SM}}}
\newcommand{\gmtwo}{(g-2)_{\mu}}
\newcommand{\BR}{BR}
\newcommand{\brbsgamma}{\BR(\overline{B}\rightarrow X_s\gamma)}
\newcommand{\cl}{\text{CL}}

\newcommand\etal{{\it {et al.}}}

\begin{document}

\begin{flushright} FTUV-10-0126, IFIC/10-03\\
PNUTP-10-A03
\end{flushright}
\vspace{.2cm}

\centerline{
\Large{\textbf{The degenerate gravitino scenario}}}
\vspace{.5cm}
\centerline{\large{Lotfi Boubekeur$^{\rm a,b}$, Ki Young Choi$^{\rm c}$, Roberto Ruiz de Austri $^{\rm b}$, and Oscar Vives $^{\rm a,b}$}}
\begin{center}
\small{
\textit{$^{\rm a}$ Departament de F\'isica Te\`orica,
Universitat de Val\`encia, E-46100, Burjassot, Spain.}}

\vspace{.2cm}
\small{\textit{$^{\rm b}$ Instituto de F\'isica Corpuscular (IFIC),
Universitat de Val\`encia-CSIC,\\ Edificio de Institutos de Paterna, Apt. 22085, E-46071, Valencia, Spain.}}

\small{
\textit{$^{\rm c}$  Department of Physics, Pusan National University, 
  Busan 609-735, Korea.}}

\vspace{.2cm}
\end{center}

\vspace{.8cm}

\hrule \vspace{0.3cm}
\noindent 
\small{\textbf{Abstract}} 
\\[0.3cm]
In this work, we explore the ``degenerate gravitino'' scenario where the mass difference between the gravitino and the lightest MSSM particle is much smaller
than the gravitino mass itself. In this case, the energy released in the decay of the
next to lightest sypersymmetric particle (NLSP) is reduced. Consequently 
the cosmological and astrophysical constraints on the gravitino abundance, and hence on the reheating temperature, become softer than in the 
usual case. On the other hand, such small mass splittings generically imply a 
much longer lifetime for the NLSP. We find that, in the constrained MSSM (CMSSM), for neutralino LSP or NLSP, 
reheating temperatures compatible with thermal leptogenesis are reached for 
small splittings of order $10^{-2}$ GeV. While for stau NLSP, temperatures of $\treh \simeq 4 \times 10^9 \gev$ can be obtained even for splittings of order of tens of GeVs. This ``degenerate gravitino'' scenario offers a possible way out to the gravitino problem for 
thermal leptogenesis in supersymmetric theories.
\vspace*{0.3cm}
\noindent\hrule \vspace{0.3cm}

\section{Introduction}

The existence of long-lived massive particles is always welcome in
theories beyond the Standard Model (SM) if their lifetime is longer
than the age of the universe (or completely stable) as they provide a
candidate for the cold Dark Matter (DM) component of the universe to
account for the abundance, $\Omega_{\rm CDM} h^2 \simeq 0.11$,
inferred by cosmological observations. However, long-lived particles
decaying after Big Bang Nucleosynthesis (BBN) place the theory in a
very difficult  situation when confronted with cosmological and
astrophysical observations\footnote{Decay products of particles with 
lifetimes as long as $10^{27}$ sec., see e.g. \cite{Picciotto:2004rp,Hooper:2004qf,Yuksel:2007dr}, can still
  produce observable signatures at present. Particles with longer
  lifetimes are  only constrained from the measured relic
  density.}. The very successful predictions of standard BBN are
spoiled by the energetic products of the decay that can dissociate the
produced light elements for lifetimes from $10^2$ to $10^{10}$
seconds. If the lifetime is between $10^{10}$ and $10^{13}$ seconds,
very stringent constraints come from the shape of the Cosmic Microwave
Background (CMB) spectrum. For longer lifetimes, the emitted photons
can reach us today and be observed as part of the Diffuse
Extragalactic Background RAdiation (DEBRA). These observations
restrict strongly the released energy in the decays of the long-lived
particle at different times and hence its abundance, mass and
lifetime.

The paradigm of such long-lived particle conflicting with cosmological
and astrophysical observations is the gravitino in supersymmetric
theories.  The gravitino is the spin $3/2$ supersymmetric partner of
the graviton and the couplings of the gravitino to ordinary matter are
gravitational couplings suppressed by the Planck mass, $M_{\rm Pl}=(8
\pi G_N)^{-1/2}\simeq 2.4\times 10^{18}\gev$.  Such small couplings
make the typical gravitino lifetime (with electroweak scale masses and
neglecting the masses of the decay products) of the order of $10^8$
seconds. Therefore, they decay after BBN and the gravitino abundance
at BBN is severely constrained \cite{Moroi:1995fs}. This has very
important consequences in the phenomenology of this model.   In fact,
these BBN constraints forbid reheating temperatures larger than $10^6$
or $10^7$ GeV, which precludes thermal leptogenesis from generating a
large enough baryon asymmetry.

It is interesting to check whether it is possible to evade these
stringent bounds. However, given that the gravitino couplings are
completely fixed by supergravity, it is very difficult to change the
gravitino production or its decay. The only ``free'' parameter available 
(from an effective theory point of view) is the gravitino mass
itself and similarly the masses of the decay products.  The gravitino
gets a mass after supersymmetry breaking and the different
soft-breaking terms receive a contribution proportional to the
gravitino mass. If supergravity is the mechanism of mediation of the
supersymmetry breaking from the hidden sector to the visible sector,
we can expect the SUSY masses and the gravitino to be of the same
order\footnote{In principle, in gauge mediation or anomaly mediation models, constraints from gravitino decays can be evaded as gravitino mass can be, respectively, much smaller or much larger than typical SUSY masses}.
However, most phenomenological analyses of BBN constraints assume that
the released energy in the thermal plasma is of the same order as the
gravitino mass itself. Although this is correct in most of the cases
where the mass difference between the gravitino and the SUSY particles
(typically the lightest MSSM supersymmetric particle (MLSP)) is sizeable, we
can ask what happens if the gravitino and the MSSM LSP masses are much
closer, $\Delta M = \mg - m_{\rm MLSP} \ll \mg$.  This has two different
consequences, first it is evident that the released energy in the
thermal plasma is much smaller and this can help to relax the previous
bounds. On the other hand, this small mass difference increases the
gravitino lifetime and other constraints as the CMB spectrum or
diffuse gamma rays come into play.

In this work, we will analyze this scenario that we call ``degenerate
gravitino'' scenario, including both the case where the gravitino is
the NLSP decaying to the MSSM LSP and the case where the gravitino
itself is the LSP and all SUSY particles decay into the gravitino.  In
the following, we define $\delta$ as the degree of degeneracy between
NLSP and LSP 
\begin{equation} 
\delta\equiv \frac{\mnlsp-\mlsp}{\mlsp} =  
\frac{\Delta M}{\mlsp}
\label{delta}
\end{equation}

In the degenerate gravitino scenario, the total dark matter abundance
has two sources: one is the usual LSP component from thermal
production and decoupling and a second one from the NLSP non-thermal
decays to LSP and SM particles.  The sum of both components should 
reproduce the observed CDM relic density.  The phenomenology of the
degenerate gravitino scenario depends on both the identity of the
lightest MSSM supersymmetric particle and on whether the gravitino is
the LSP or the NLSP. The BBN, CMB and DEBRA bounds apply to the NLSP
abundance while the LSP abundance is only constrained by the total
dark matter abundance.  In most cases, the lightest MSSM supersymmetric particle can
be either the lightest neutralino or the lightest stau. Given that
dark matter can not be a charged particle, we are left with three
possibilities: gravitino LSP with neutralino or stau NLSP and gravitino
NLSP with neutralino LSP. We will see that, for lifetimes smaller than the age of the universe, the strongest constraints
on the reheating temperature arise in the case of gravitino NLSP,
where all the previous constraints apply to the gravitino abundance
and thus on the reheating temperature. On the other hand, when the
gravitino is the LSP, it is possible to reach reheating temperatures
compatible with thermal leptogenesis provided that the NLSP abundance
at decoupling is sufficiently suppressed. Clearly, if the NLSP lifetime is 
much longer than the age of the universe, corresponding to small enough $\delta$, 
the maximal reheating temperatures consistent with the observed DM abundance
($\treh \simeq 4.1 \times 10^9$ GeV) can be reached.

In this work, we
will reanalyze the different constraints on the energy injected by
NLSP decays and the reheating temperature in terms of
this parameter $\delta$ in the three above mentioned cases. 
In the next section, we will present the constraints on the
energy release for different lifetimes of the NLSP in a model
independent way.  In Section \ref{sec3}  we apply these
constraints to the gravitino case where the energy release and the
lifetime are related. In section \ref{sec4} we analyse the case of
the CMSSM looking for the new constraints on the reheating temperature
in the degenerate gravitino scenario. Finally in section
\ref{sec:conclusions} we present our conclusions.


\section{Model-independent bounds}
\label{sec1}

If the NLSP lifetime is smaller than the age of the Universe, the
energetic decay products can significantly affect cosmology. Even for 
lifetimes slightly larger than the age of the universe, observation of 
diffuse gamma rays can constraint this scenario. 
  In this section,  we consider the model-independent constraints on the
degenerate NLSP-LSP scenario. As we will see, depending on the NLSP
lifetime, these constraints originate from BBN observations, CMB 
spectral distortion or searches of diffuse gamma rays. There are additional 
constraints if the NLSP is charged.  We focus on the case that the NLSP and LSP
masses are nearly degenerate i.e. $\delta\ll 1$ in \eq{delta}.

\subsection{Relic abundance constraint}
The first constraint that should be considered is that the total relic density of cold dark matter must match the observed value. We will consider the situation where only the LSP and the NLSP are relevant; that is, heavier MSSM particles have already decayed to LSP/NLSP before BBN takes place. In general, the final abundance of the LSP will have two components. The first one is the thermal abundance, $\Omega^{\rm TP}_{\rm LSP}$, which comes from thermal  processes occurring in the plasma like scatterings and freeze-out.  The second one is the non-thermal component, $\Omega^{\rm NTP}_{\rm LSP}$, that includes the contribution from LSP particles produced in NLSP decay.  Therefore, the total cold dark matter relic density, will be \footnote{Recall that we are assuming that the NLSP lifetime is smaller than the age of the Universe. If it were not the case, Eq.~(\ref{abundance}) is still valid as we are considering $m_{\rm NLSP}\simeq m_{\rm LSP}$.}
\begin{equation}
\Omega_{\rm CDM}\,h^2 = \Omega^{\rm TP}_{\rm LSP}\,h^2+ 
\frac{1}{1+\delta} \, \Omega^{\rm TP}_{\rm NLSP}\,h^2\,,
\label{abundance}
\end{equation}
where the last term stands for $\Omega^{\rm NTP}_{\rm LSP}$ and we have used \eq{delta}. Assuming no late entropy release, Eq.~(\ref{abundance}) represents the present amount of dark matter, which should be equated with the observed value \cite{Komatsu:2008hk}
\be
\Omega_{\rm WMAP}\,h^2= 0.1131\pm0.0034
\label{wmap}
\ee
It is useful to define the new parameter 
\be
\omega
\equiv\frac{\ynlsp}{Y_{\rm CDM}}\;,
\label{omega}
\ee
which quantifies the amount of present cold dark matter coming from the NLSP decay. In this equation $\ynlsp$ refers to the NLSP yield just before its decay \footnote{Here the yield of a species $i$ is defined as the ratio of the number density $n_i$ to entropy, $Y_i\equiv n_i/s$. Recall also that the yield and the relic density of a massive particle species $i$ are related through $Y_i \simeq 4.1\times 10^{-12} \(100\gev/ m_i\)\(\Omega_i\, h^2/ 0.11\)
$.}. Combining \eq{abundance} and \eq{omega} we get that
\be
\omega=1-\frac{\Omega^{\rm TP}_{\rm LSP}\,h^2}{\Omega_{\rm WMAP}\, h^2}\, ,
\ee
which implies that $\omega\le 1$ independently of $\delta$.

\subsection{BBN constraints}
At temperatures of order $T\sim 1$ MeV, the light nuclei are synthesized in the primordial plasma. These temperatures corresponds to times between 1 sec and $10^3$ sec. The obtained abundances in standard BBN calculations are in striking agreement with observation \footnote{There are possible discrepancies in Lithium abundances, which  might be  
explained using Gravitino dark matter ~\cite{Jedamzik:2005dh,Bailly:2008yy}.
}. 
However, the injection of energetic particles in the primordial plasma at BBN or later can disrupt the standard
BBN processes~\cite{Cyburt:2002uv,Jedamzik:2004er,kkm04,Jedamzik:2006xz}, leading to a disagreement between theory and observation.  
Thus, any additional particle decaying at BBN or later is subject to the strong constraints from light nuclei abundances.  
Usually, the energy of injected particles is assumed to be of the same order of the LSP/NLSP mass and stringent bounds were derived for 
various scenarios from BBN~\cite{Ellis:1984eq,fst04-sugra,Ellis:2003dn,Roszkowski:2004jd,Kohri:2005wn,Cerdeno:2005eu,
Steffen:2006hw,Pradler:2006qh,Kawasaki:2008qe,Bailly:2009pe}.
These constraints are obtained by solving the full set of Boltzmann equations for BBN with a late-decaying particle. This requires the detailed study of the spectrum of decay products with all the relevant
nuclear cross sections~\cite{Jedamzik:2004er,kkm04,Jedamzik:2006xz}.  
The constraints apply to the released electromagnetic or hadronic energy, parametrised by $\xi_{i}$ defined as \cite{fst04-sugra} 
\dis{
\xi_{i}\equiv E_{i}\,B_{i}\,\ynlsp \label{xiem}\;, 
}
where $E_i$ is the released energy per decay with $i=$ em for electromagnetic decays and $i=$ had for the hadronic ones and $B_i$ stands for the respective branching ratios. The constraints from hadronic processes are important when the lifetime is relatively
short $\tau_{\rm NLSP}\lesssim 10^7$ sec \cite{kkm04,Kohri:2005wn}. The typical lifetimes considered in this work are 
larger that $10^7$ sec and therefore, in the following, we will consider only the BBN constraints on $\xi_{\rm em}$.

In the 2-body electromagnetic decay, the corresponding released energy is 
\dis{
E_{\rm em}= \frac{\mnlsp^2-\mlsp^2}{2\mnlsp}\, ,
}
where we assumed that the visible particle mass is negligible. In the degenerate mass limit, $\delta \ll 1$, this reduces to $E_{\rm em}\simeq \mlsp \,\delta$. 
Notice that, as emphasized before, $E_{\rm em}$ is much smaller than the usually\footnote{Small mass differences has been considered in \cite{Bailly:2009pe,Cembranos:2007fj} and for completely different motivations than ours.} assumed value $\mnlsp/2$.

Given the above discussion, for our purposes, we can apply the constraints on $\xi_{\rm em}$ from Ref.~\cite{Jedamzik:2006xz} directly to our scenario. Then, the constraint on  $\xi_{\rm em}$ reads
\dis{
\xi_{\rm em}
 \simeq
4.1 \times 10^{-10} \gev \bfrac{\abundwmap}{0.11}\, \omega\, B_{\rm em}\,\delta 
< \xi_{\textrm {upper limit}}\,, \label{xiBBN}
}
where the right hand side in the above equation can be read off from the upper limit derived in \cite{Jedamzik:2006xz} for $B_{\rm had}=0$.  

\begin{figure}[!t]
\begin{center}
\vspace{-1cm}
\includegraphics[width=0.6\textwidth]{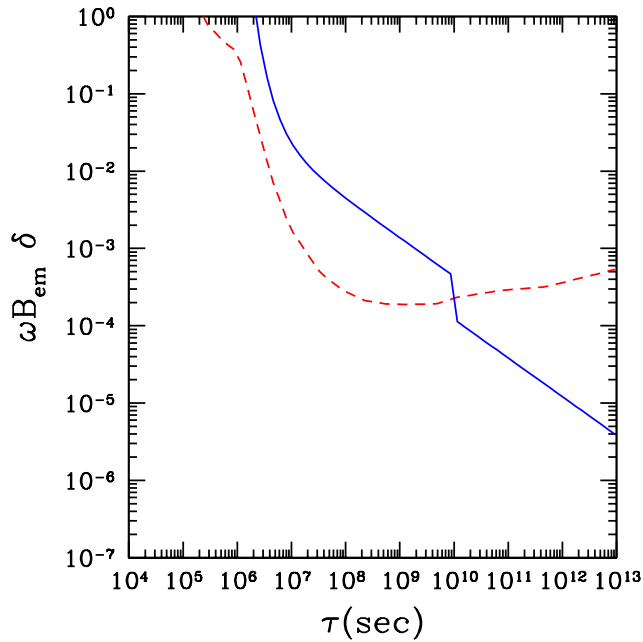} 
\vspace{-0.7cm}
 \caption{ \small Constraints on the 
combined parameters $\omega B_{\rm em}\delta$ versus lifetime
of NLSP using the result of~\cite{Jedamzik:2006xz} with $B_h=0$. 
The red (dashed) line come from em BBN constraints and blue (solid) line 
from CMB distortion as explained in the text. 
Regions above the lines are excluded.} 
\label{fig:BBNCMB}
\end{center}

\end{figure}

For general values, the upper limit on the product $\omega\, B_{\rm em}\,\delta$ from BBN is shown in Figure~\ref{fig:BBNCMB} with red (dashed) line.
From this figure, we can see that for lifetimes between $10^7$ and $10^{10}$ sec, this constraint requires $\omega B_{\rm em}\delta$ to be between $10^{-3}$ and $10^{-4}$. Notice that BBN constraints apply equally to the case of charged NLSP decaying to gravitino and electromagnetic showers.

\subsection{CMB spectral distortion}
In addition to BBN constrains, for long lifetimes $\tau_{\rm NLSP} \gsim 10^7$ sec., there are strong bounds from the shape of the CMB black-body spectrum. As pointed out in~\cite{Ellis:1984eq}, the late injection of 
electromagnetic energy may distort the frequency dependence of the CMB
spectrum from its observed blackbody shape. At late times of
interest in our scenario, energetic photons from
NLSP decays lose energy through processes such as $\gamma
e^-\rightarrow \gamma e^-$, but photon number remains conserved since
other processes, like double Compton scatterings and thermal
bremsstrahlung, become inefficient. As a result, the spectrum follows
the Bose--Einstein distribution function
\begin{equation}
f_\gamma(E)=\frac{1}{e^{E/(kT)+\mu}-1},
\label{eq:cmbbedist}
\end{equation}
where $\mu$ here denotes the chemical potential. 
Then, the chemical potential of the distorted CMB spectrum has to satisfy the constraint, $|\mu|< 9\times 10^{-5}$~\cite{mubound}, which, for decay lifetimes $\taunlsp \lsim 8.8\times10^{9} \sec$, translates into an upper bound on
the released energy from NLSP decay, $\xi_{\rm em}$ defined in \eq{xiem}
\cite{Hu:1993gc,Roszkowski:2004jd},
\dis{
\xi_{\rm em} <& 1.59\times 10^{-8} \,
  e^{(\tau_{dC}/\taunlsp)^{5/4}}\left({1 \sec \over \taunlsp} \right)^{1/2}\gev
\label{xiCMB},
}
where $\tau_{dC}\simeq 6.085\times 10^6$ sec.

For longer lifetimes ($\taunlsp \gsim 8.8\times 10^{9} \sec$),  
the spectral distortions in
the CMB spectrum can be described in terms of the integral of the fractional
contributions to the energy $\epsilon$ of the CMB per comoving volume during
decay through the Compton $y$ parameter, $4y=\delta\epsilon/\epsilon$, given by
\begin{eqnarray}
\frac{\delta\epsilon}{\epsilon}=7.04\times\frac{1}{T(t_{\rm eff})}\xi_{\rm em},
\label{eq:cmbypar}
\end{eqnarray}
where $T(t)$ is the CMB temperature and $t_{\rm
eff}=[\Gamma(1-\beta)]^{1/\beta}\taunlsp$, for a time--temperature relation
$T\propto t^{-\beta}$, with $\Gamma$ the usual Gamma function.
In the radiation dominated era in the early Universe, for $T
< 0.1 \mev$, 
\begin{eqnarray}
T=1.15\times 10^{-3}\left(\frac{t}{1\sec}\right)^{-1/2} \gev, 
\end{eqnarray}
which gives
$\beta=1/2$.  Thus $t_{eff}= [\Gamma(1/2)]^2 \taunlsp=\pi\taunlsp$.

The observational limit $|y|<1.2\times 10^{-5}$~\cite{Hagiwara:2002fs} 
gives the constraint
\dis{
\xi_{\rm em}\lesssim 4.42\times 10^{-9}
\gev\sqrt{\frac{1\sec}{\taunlsp}}.
\label{eq:yboundonxiem}
}
Using Eqs.~(\ref{eq:cmbypar},\ref{eq:yboundonxiem}), this CMB constraint 
is plotted in Figure~\ref{fig:BBNCMB} with blue (solid) line.
We can see that for lifetimes $\taunlsp \gtrsim 10^{10}$ sec. the CMB 
constraint becomes more stringent than the BBN ones and sets the constraint on $\xi_{\rm em}$ until recombination time. 
Similarly to the BBN, CMB constraints apply also to charged NLSP decaying to gravitino and electromagnetic showers.

\subsection{Diffuse Gamma-ray observation}

After recombination, at the cosmic time around $10^{13}$ sec,
the number density of free electrons drops quickly and the photons are almost
free from the interactions. 
Therefore the photons from the decaying particles can reach us now and
contribute to the Cosmic Gamma-ray Background (CGB).
The observed CGB highly constrains any extra contribution
 including the photons from late decaying particles
\footnote{Under appropriate conditions, the late decays of WIMPs to gravitinos 
and MeV photons
may explain the MeV CGB anomalies~\cite{Cembranos:2007fj}.}.

The present photon flux from two-body decay can be written as
\dis{
\frac{d\Phi}{dE_\gamma}=\frac{c}{4\pi}\int^{t_0}_{t_i}\frac{dt}{\taunlsp}\frac{\rho_c\,\Omega_{\rm WMAP}\,\omega\,
B_{\rm em}}{\mnlsp}\,e^{-t/\taunlsp}\delta(E_\gamma
-a E_{\rm em}),
}
where $E_{\rm em}\simeq \mlsp \delta$ is the energy of the photon at
production, $\taunlsp$ and $\mnlsp$ are the lifetime and mass of NLSP,
$\rho_c=3H_0^2/8\pi G_N=8.0992h^2\times 10^{-47}\gev^4$ and $a=a(t)$ is
the time-dependent scale factor with $a(t_0)=1$ at present time $t_0$.
The delta function can be integrated using the formula
$\delta(f(t))=\delta(t-t_a)/|f'(t=t_a)|$ where $t_a$ is the solution which 
satisfies $E_\gamma=a(t_a) E_{\rm em}$. After integration, the flux reads
\dis{
\frac{d\Phi}{dE_\gamma}=\frac{c}{4\pi}\frac{\rho_c\,\Omega_{\rm WMAP}\,\omega \,B_{\rm em}}{\mnlsp\,\taunlsp}\,\frac{e^{-t_a/\taunlsp}}{E_\gamma
H(E_\gamma/E_{\rm em})}\Theta(E_{\rm em}-E_\gamma),
}
where the $\Theta$ function simply cuts energies larger than the initial energy
and  each observed photon with energy $E_\gamma$ is produced at a time $t_a$ which satisfies
$a(t_a)=E_\gamma/E_{\rm em}$. Assuming the dark energy is a cosmological constant,
this function $t_a(a)$ is given by~\cite{Cembranos:2007fj}
\dis{
t_a\equiv t (a=E_\gamma/E_{\rm em})=\frac{2\log [(\sqrt{\Omega_\Lambda a^3} +\sqrt{ \Omega_M  +\Omega_\Lambda a^3})/\sqrt{\Omega_M}  ]}{3H_0 \sqrt{\Omega_\Lambda}},
}
where $H(a)=H_0\sqrt{\Omega_M a^{-3} + \Omega_\Lambda}$.

Taking into account that $c/H_0=1.3\times 10^{28} \cm$,
$\rho_c=5.6\times 10^{-6} \gev/\cm^3$, $H_0=70 \km\,\sec^{-1}\, \mpc^{-1}$,
$\Omega_\Lambda=0.7$, $\Omega_{M}=\Omega_{\rm CDM}+\Omega_B=0.3$ and
$\Omega_{\rm CDM}=\Omega_{\rm WMAP}=0.25$, we find that the flux is
\dis{
\frac{d\Phi}{dE_\gamma}=1.37\times 10^{21}~ \gev \cm^{-2} \frac{\omega B_{\rm em}}{\mnlsp \taunlsp}
\frac{e^{-t_a/\taunlsp}}{E_\gamma \sqrt{0.7+0.3(E_{\rm em}/E_\gamma)^3 } },
\label{CGBspec}
}
with
\dis{
t_a=3.51\times 10^{17} \sec \log[\sqrt{2.33a^3}+\sqrt{1+2.33 a^3}].
}

This differential flux must be compared to the observation of
diffuse gamma ray flux. At each photon energy, we must require that this 
flux is smaller than the flux, $E_\gamma^2 \frac{d \Phi}{d E_{\gamma}}$, 
observed by SPI, COMPTEL and EGRET \cite{Yuksel:2007dr}.
Notice that, unlike the BBN and CMB constraints where we can find a bound on $\xi_{\rm em}$ for a given value of $\taunlsp$, now we need to specify both $\taunlsp$ and $E_\gamma$ to obtain a bound on $\omega B_{\rm em} \delta$. 

On the other hand, if the lifetime of NLSP is longer than the age of Universe,
the line spectrum from the galactic center can be observed without
cosmological redshift and this provides a further constraint on the emitted radiation from NLSP decays. For this we apply the bounds from 
Ref.~\cite{Yuksel:2007dr}.
Taking into account of the exponential decay of NLSP we have
\dis{
\frac{\rho_{sc}\omega B_{\rm em}}{4\pi\mnlsp \taunlsp}e^{-t_0/\taunlsp} \zeta_{lim} < \mathcal{F}(E_\gamma=E_{\rm em}),
}
where $\rho_{sc}=0.3 \gev \cm^{-3}$ is the dark matter density at the solar
distance from the Galactic center, $R_{sc}=8.5\kpc$, $\zeta_{lim}$ is a
dimensionless integral of the line-of-sight intensity in the galactic center
which ranges between $0.5 -1.5$ for various dark matter halo
profiles~\cite{Yuksel:2007dr}. The function $\mathcal{F}$ is given in Figure 2 
of Ref.~\cite{Yuksel:2007dr}. Once again, as in 
the case of the CGB constraint in \eq{CGBspec}, this constraint depends on the photon energy and therefore we need to specify both $\taunlsp$ and $E_\gamma$ to obtain bound on $\omega B_{\rm em} \delta$. Thus, there is no simple analog of Figure~\ref{fig:BBNCMB} for a 
constraint on $\xi_{\rm em}$ from Diffuse gamma rays observations. Moreover, the constraints on  $\xi_{\rm em}$ from the galactic center gamma rays 
are of the same order of magnitude (although slightly stronger) as the ones from the diffuse extragalactic emission. Therefore, for 
simplicity, we will only consider the diffuse gamma rays constraints, which apply to a broader range of energy.

\subsection{Catalyzed BBN}

Heavy long-lived negatively charged particles, $X^-$, present during BBN can
bind with light nuclei modifying standard BBN reactions.  These
catalyzed reactions, called CBBN, result in a change of light element
abundances, and in particular lead to the overproduction of
$^{6}\textrm{Li}$ through CBBN reactions ~\cite{CBBN}.
 For lifetimes longer than $5\times 10^3 \sec$, the observed light-nuclei abundances result on a constraint on the abundance of the charged relic, $Y_{X^-} < 2\times10^{-16}$~\cite{Hamaguchi:2007mp,Pradler:2007is}. 
However, taking more conservative $^{6}\textrm{Li}/ ^{7}\textrm{Li}$ constraints, it is possible to relax slightly the previous bound to $Y_{X^-} < 10^{-14}$--$10^{-15}$ ~\cite{Jedamzik:2007qk,Jedamzik:2009uy}.

Using \eq{omega}, we can translate the constraint on the yield of charged NLSP
from catalyzed BBN to a bound on $\omega$
\dis{
\omega 
\lesssim 
2.44\times10^{-3}\bfrac{\mlsp}{100\gev}\bfrac{Y_{\textrm{CBBN}}}{10^{-14}}
\label{CBBNomega}
}
where we used $\abundwmap=0.11$ and $Y_{\textrm{CBBN}}$ is the maximum value 
allowed from catalyzed BBN for lifetimes larger than $10^5\sec$. 
As can be see from this equation, the catalyzed BBN bound is very stringent 
and indeed, it is very difficult to obtain such small yield at freeze-out in the MSSM \cite{Berger:2008ti}. However, it is still possible to find small 
allowed regions in the MSSM, or even in the CMSSM,  where the $\stau$ is the NLSP with large $\tan \beta$ \cite{Ratz:2008qh,Pradler:2008qc}  with relaxed $^{6}\textrm{Li}/ ^{7}\textrm{Li}$ bounds~\cite{Bailly:2009pe}.

\section{Bounds on the degenerate gravitino scenario}
\label{sec3}

In this section, we will apply the model-independent bounds derived in the last section to the case of degenerate gravitino scenario. In our scenario, the gravitino can be either the LSP or the NLSP. In both cases it is thermally produced at reheating. After inflation, the Universe is reheated at a temperature 
$T_{\rm RH}$ and gravitinos are produced through thermal scatterings in the plasma \footnote{We assume that at $\treh$, the Universe is composed of a thermal bath of MSSM degrees of freedom and that gravitinos production by inflaton decay is negligible \cite{Nilles:2001ry}.}. Their resulting relic density is linear in the reheating temperature and it is given by \cite{Bolz:2000fu}\footnote{Taking into account the result of Ref.~\cite{Rychkov:2007uq} the gravitino relic density would roughly increase a factor 2.}
\begin{equation}
\Omega^{\rm TP}_{3/2}\,h^2\simeq 0.27 \(\frac{T_{\rm RH}}{10^{10}\gev}\)\(\frac{100\gev}{\mg}\)\(\frac{M_3}{\tev}\)^2\,,
\label{omega1}
\end{equation}
where $M_3$ is the gluinos mass. Assuming no late entropy release, \eq{omega} sets an absolute bound on the reheating temperature, i.e. $\Omega^{\rm TP}_{3/2}\lesssim \Omega_{\rm WMAP}$, implies 
\begin{equation}
T_{\rm RH}\lesssim 4.1\times 10^9 \gev\,   \(\frac{\mg}{100\gev}\)
\(\frac{\tev}{M_3}\)^2\,. 
\label{max}
\end{equation}
This has to be compared with the minimum reheating temperature for successful thermal leptogenesis   \cite{Davidson:2002qv,Giudice:2003jh, Antusch:2006gy} $T_{\rm RH} \gtrsim 2\times  10^9 \gev$. From this equation it is clear that the reheating temperature can not reach values much above $\sim 10^{10}$ GeV.   \footnote{A possibility to increase $T_{\rm RH}$ is to consider a squeezed gaugino spectrum  reducing the gluino mass $M_3$ (see e.g. \cite{Olechowski:2009bd}).} In addition to this constraint, we have to implement the constraints on the released energy considered in the last section.

The relevant particles in the analysis of NLSP decays in the degenerate scenario are the gravitino and the lightest MSSM particle, which can be either the neutralino or the stau\footnote{Sneutrinos LSP are marginally  allowed \cite{Arina:2007tm} due to their too large direct detection cross-sections. We do not consider them in our analysis.}.
We have two different situations depending on the particle nearly degenerate with the gravitino, namely gravitino-neutralino and gravitino-stau degeneracy. In each case the gravitino can be either the LSP or the NLSP.

\subsection{Gravitino-neutralino degeneracy}
As usual, in the MSSM, the lightest neutralino eigenstate $\chi^0_1$ is 
parametrised as  
$\chi^0_i = N_{i 1} (-i\widetilde{B})+ N_{i 2} (-i \widetilde{W}_3) + N_{i 3}\widetilde{H}^0_U + N_{i 4}\widetilde{H}^0_D$, where the unitary matrix $N$ defines  the composition of neutralinos in terms of the Bino, Wino and Higgsinos. Since we are considering mass splittings  that are smaller than the $Z$ mass to suppress the hadronic branching ratio, the dominant (2-body) decay channel will be  $\chi^0_1\to \gamma\,\widetilde{G}$ or  $\widetilde{G}\to\chi^0_1 \gamma$. As we will see below, the typical lifetime of NLSP is of order $10^{13}\sec \times \left( 1 \gev/ \Delta M\right)^3$. Therefore, depending on $\Delta M$ we will have to consider different constraints.   

\subsubsection{Gravitino LSP}

In this case, the lifetime of neutralinos is given by 
\begin{equation}
\tau_{\chi}\simeq \frac{1.78\,\times 10^{13}\sec}{|N_{11}\cos\theta_W+N_{12}\sin\theta_W |^2} \left(\trac{ 1 \gev}{ \Delta M}\right)^3\,.
\end{equation}

\begin{figure}[!t]
  \begin{center}
\includegraphics[width=0.6\textwidth]{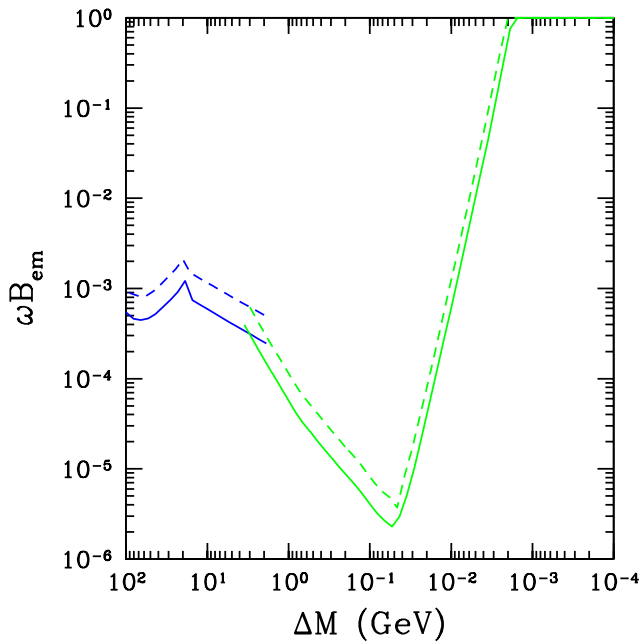} 
  \end{center}
\vspace{-1cm}
 \caption{\small Constraints on the 
combined parameters $\omega B_{\rm em}$ versus $\Delta M$
for given mass of neutralino NLSP (for $N_{11} \simeq 1$) with Gravitino LSP. 
Solid line is for $\mnlsp=100\gev$ and dashed line for $\mnlsp=200\gev$. 
The blue lines for $\Delta M > 2 \gev$ come from BBN and CMB constraints
and the green lines for $\Delta M< 2 \gev$ come from the diffuse gamma ray observations.
Regions above the lines are excluded. } 
\label{fig:combinedneutr}
\end{figure}

Notice that, in the limit of $\Delta M \ll m_{3/2}$, the lifetime depends only on the mass splitting $\delta\times\mg=\Delta M$, but not on the overall mass scale $\mg$ \cite{Moroi:1995fs,fst04-sugra}. 
If the gravitino is the LSP,  the neutralino will decay into gravitino and photon and there are strong constraints on the released energy $\xi_{\rm em} = \omega\, B_{\rm em} \,\delta$. Different constraints will apply depending on the lifetime which, in turn is fixed by $\Delta M$. 
\begin{itemize}
\item For $10 \gev \lesssim 
\Delta M \lesssim 90 \gev$, the relevant constraint is BBN and from Fig.~\ref{fig:BBNCMB} we have roughly $\omega \delta \lesssim 10^{-3}$. 
\item For smaller splittings, from $1 \gev \lesssim \Delta M \lesssim 10 \gev$,  the constraints from CMB spectrum are much stronger than the BBN ones and $\omega \delta$ is between $10^{-4}$ and $10^{-6}$  
\item For $ 30 \mev \lesssim \Delta M \lesssim 1 \gev$ we have to take into account the diffuse gamma ray observations. Although the constrains on $\omega \delta$ depend on $m_{3/2}$,  for $m_{3/2}\simeq 100$ GeV, typical values range from $\omega \delta = 10^{-6}$ to  $\omega \delta = 10^{-9}$.  
\item Mass differences from $ 2 \mev \lesssim \Delta M \lesssim 30 \mev$ correspond to NLSP with lifetimes $  2 \times 10^{21} \sec \gtrsim \tau_{\rm NLSP} \gtrsim  5 \times 10^{17} \sec$. These NLSP are already decaying at present, therefore diffuse gamma ray observations constrain their abundance. In this case  $\omega \delta$ ranges from $10^{-9}$ to $10^{-5}$.
\item Finally, for smaller $\Delta M$, the neutralino is still present in the 
universe and it is completely stable for practical purposes. The only constraint comes from the WMAP measurement of the dark matter abundance.  
\end{itemize}
These constraints are summarized in Figure~\ref{fig:combinedneutr}. We can see that for a given $\Delta M$ the constraints on $\omega$ are very strong for $\Delta M > 30 \mev$, and, as we will see, it is difficult to reach such small neutralino abundances at decoupling in the MSSM. For smaller $\Delta M<2 \mev$,  corresponding to neutralino lifetimes longer than the age of the universe, $\omega \simeq 1$ is evidently allowed. For $2 \mev<\Delta M <30\mev$, $\omega$ ranges from 1 to $10^{-6}$.  

The only constraint on the gravitino abundance comes from \eq{abundance} and the dark matter abundance measured by WMAP. Given $\omega$, we can use Eq.~(\ref{delta}) and Eq.~(\ref{omega}) to calculate the required reheating 
temperature
\begin{eqnarray}
\treh&=& 4.1\times10^{9}\gev\left(\frac{\mg}{100\gev} \right)\left(\frac{1\tev}{M_3}\right)^2 \left(1-\omega\right).
\label{treh_GLSP}
\end{eqnarray}

From Eq.~(\ref{treh_GLSP}), we see that, provided $\omega\ll 1$, one gets the maximal allowed reheating temperature $\treh = 4.1 \times 10^9$ GeV.

\subsubsection{Gravitino NLSP}

If the gravitino is the NLSP, the dominant decay channel is $\widetilde{G}\to \gamma\,\chi^0_1$ and its lifetime is given by 
\begin{equation}
\tau_{3/2}\simeq \frac{3.56\,\times 10^{13}\sec}{|N_{11}\cos\theta_W+N_{12}\sin\theta_W |^2} \left(\trac{ 1 \gev}{ \Delta M}\right)^3\,,
\end{equation}
which is only a factor 2 larger than the gravitino LSP case. However, unlike the gravitino LSP case, the bounds on the released energy from this decay constrain strongly the initial thermal abundance of gravitinos. Notice that, here, $\omega$ represents the fraction of neutralinos coming from gravitino decay and therefore the initial abundance of gravitinos.

\begin{figure}[!t]
  \begin{center}
 \begin{tabular}{c c}
\includegraphics[width=0.6\textwidth]{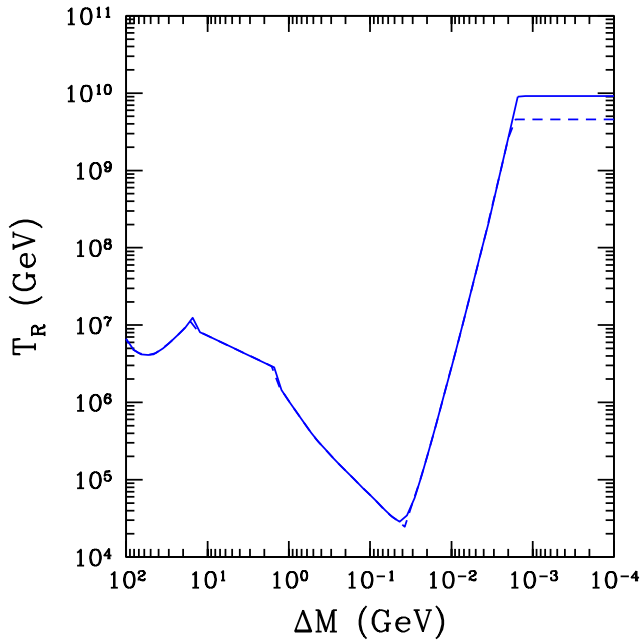} 
&
\end{tabular}
  \end{center}
\vspace{-1cm}
 \caption{\small Constraints on the maximal reheating temperature after inflation
for the Gravitino NLSP with neutralino LSP (with $N_{11} \simeq 1$) 
for $\mnlsp= 100\gev$ (solid)
and $200\gev$ (dashed). Regions above the lines are excluded. Notice that, as we can see  from \eq{treh_GNLSP}, these maximal reheating temperatures are only possible if the corresponding bounds on $\omega$ are satisfied} 
\label{fig:treh}
\end{figure}

The constraints on $\omega \delta$ for different ranges of $\Delta M$ seen in the previous section apply equally in this case. Now, given $\Delta M$ and $m_{3/2}$, which fix $\delta$, we have a direct constraint on the gravitino thermal abundance, and hence on $\omega$. As we know, the gravitino thermal abundance is directly proportional to $\treh$ which in this case can be written
\dis{
\treh\simeq 4.1\times10^{9}\gev\left(\frac{\mg}{100\gev} \right)\left(\frac{1\tev}{M_3}\right)^2 \omega \left(\frac{1}{1+\delta}\right) .
\label{treh_GNLSP}
}
From this equation, we see that, opposite to the case of gravitino LSP, in order to maximize  the reheating temperature, one needs $\omega$ as large as possible. The constraints on $\treh$ as a function of $\delta$ are shown in Figure~\ref{fig:treh}. Given the strong constraints on $\omega$ for the different $\delta$s,
$\treh$ is considerably smaller than the maximal value allowed by WMAP. The only exception to this are the cases with very small $\Delta M \lesssim 2 \times 10^{-3}$ GeV, corresponding to the NLSP still present as a dark matter component, where $\omega\lesssim 1$ and the LSP abundance is smaller that the observed $\Omega_{\rm WMAP}\,h^2$.

\subsection{ Gravitino-stau degeneracy}

In the MSSM, the lightest stau state $\stau_1$ can also be the LSP. However this situation is usually discarded as staus  cannot play the role of cold dark matter. The remaining possibility is then that $\stau_1$ is the NLSP with the gravitino as the LSP. The dominant decay channel in this case is  $\stau_1 \to \tau\,\widetilde{G}$ if the mass-difference is larger than the tau mass, i.e. $\Delta M \geq 1.77 \gev$~~\footnote{Notice that for $\Delta M \leq 1.77 \gev$ the two body flavour-conserving channel is closed and the stau can decay only through lepton flavour violating channels, $\stau_1 \to \mu\,\widetilde{G}$ or  $\stau_1 \to e\,\widetilde{G}$, where the lifetime would be inversely proportional to the lepton-flavour violating coupling \cite{Masiero:2004js,Kaneko:2008re}: $\tau_{\stau_1}\simeq 2\times 10^{14}\sec |\delta^{\rm LFV}_{\tau i}|^{-2}$ for $\mg =100 \gev$ and $\Delta M = 2 \gev$. This means that, if these flavour violating couplings, the so-called Mass Insertions, are sizable, $\delta^{\rm LFV}_{\tau i}\geq 0.03$, the stau could decay before the present time. However, we do not consider this possibility in this paper. }. In this case, the lifetime of the stau NLSP is then given by 
\begin{equation}
\tau_{\stau_1}\simeq  6.69\times 10^{15}\sec \left(\trac{ m_{3/2}}{300 \gev}\right) \left(\trac{2 \gev }{ \Delta M}\right)^4   \left(\trac{0.21}{ 1 - m_\tau^2/(\Delta M)^2}\right)^{3/2} \,.
\end{equation}

Notice that now, the stau lifetime, in contrast with the degenerate
gravitino-neutralino case, does depend on the overall mass scale $\mg$. 
Moreover for a mass difference similar to the tau mass, $\Delta M\simeq 1.78 \gev$, the stau lifetime is equal to the age of the universe.   

In principle, the stau has to satisfy similar constraints as in
the degenerate gravitino-neutralino case with the exception of diffuse
gamma ray constraints. Notice that the stau does not decay directly to
photons and therefore does not contribute to the gamma ray
background. Despite this fact, all the decays of the stau produce
electromagnetic cascades that affect both BBN and CMB observables.

Nevertheless, the main difference with the neutralino case is that
stau, being charged, can form bound states with light elements and
affect BBN predictions \cite{Pradler:2007is, Jedamzik:2007qk}. Then,
from \eq{CBBNomega} we obtain a strong constraint on $\omega$ which
for $Y_{\rm CBBN}= 10^{-15}$ is $\omega < 7.32 \times 10^{-4}
m_\stau/(300 {\rm GeV})$.

Finally, if the stau lifetime is longer than the age of the universe,
 the stau yield is very strongly bounded by
the presence of anomalously heavy Hydrogen in deep sea water \cite{Yamagata:1993jq}. In terms of $\omega$ the bound reads, $\omega \leq 2.2 \times 10^{-27}\left(\mstau/100 \gev\right)$, for stau masses between 5 GeV and 1.7 TeV.
Thus, in practice, this possibility can be completely discarded. In our analysis, 
we require that staus have already decayed at present and therefore we 
eliminate the staus with a lifetime longer than the age of the universe which
corresponds to $\Delta M \lesssim 2 \gev$.  

\begin{figure}[!t]
  \begin{center}
\includegraphics[width=0.6\textwidth]{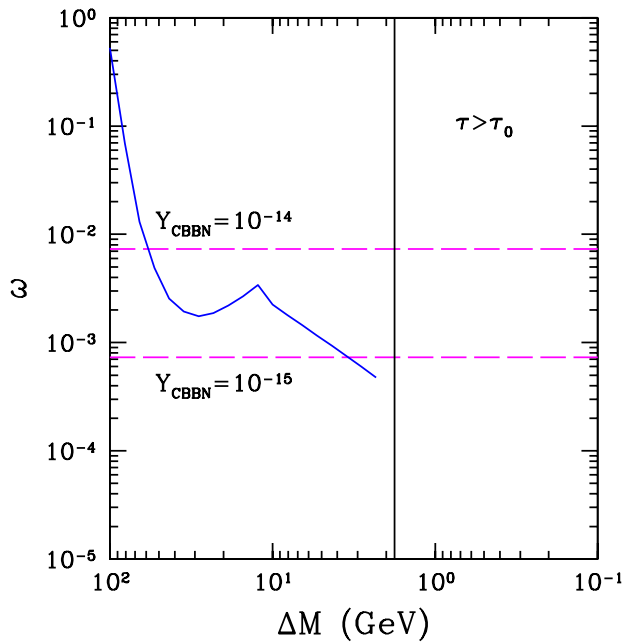}
  \end{center}
\vspace{-1cm}
 \caption{\small Constraints on $\omega$ versus $\Delta M$ with $\mg = 300 \gev$
for stau NLSP with Gravitino LSP. 
The solid blue line represents the BBN and CMB constraints. 
The vertical line at  $\Delta M \simeq 2 \gev$  corresponds to a stau lifetime equal to the age of the universe and therefore the region on the right is excluded from heavy-water searches. The dashed pink lines correspond to the CBBN constraint for
 $Y_{\rm CBBN}= 10^{-14}$ and  $Y_{\rm CBBN}= 10^{-15}$.} 
\label{fig:combinedstau}
\end{figure}

These constraints are presented in Figure~\ref{fig:combinedstau} where we see
that for $Y_{\rm CBBN}= 10^{-15}$ the CBBN is the most stringent of all the 
constraints and requires $\omega \lesssim 7\times 10^{-4}$.
This corresponds to stau relic density well below the
observed relic density, \eq{wmap},  and the thermal relic density at freeze-out 
for purely RH staus \cite{Berger:2008ti}. In order to get such small abundance, 
the staus must have a substantial LR mixing. This requires
both large $\mu$ and large $\tan\beta$ and moderate $m_{\stau_1}$
\cite{Ratz:2008qh, Pradler:2008qc}, which as we will see in section \ref{sec4}, is difficult, but still possible in the CMSSM.
Furthermore, the vertical line at $\Delta M \simeq 2 \gev$ corresponds to our requirement that all staus have already decayed at the present age.

The bound on $\treh$ here is analogous to the case of
gravitino-neutralino degeneracy with gravitino LSP. Again, the
reheating temperature is given in \eq{treh_GLSP} and given that, in this case,
the allowed points in parameter space require $\omega \lesssim
10^{-2}$, one gets the maximal reheating temperature $\treh = 4.1
\times 10^9$ GeV.

\section{The degenerate gravitino scenario in the CMSSM }

\label{sec4}

Perhaps the most appealing mechanism to transmit SUSY breaking
from a hidden sector,  where SUSY breaking occurs,  to the visible 
sector is to use gravitational interactions, that are suppressed by the Planck scale \footnote{Of course, in gravity mediated scenarios, there can be an associated cosmological moduli problem.  
In this work, as this issue is outside the focus of the paper, we do not address it and we will just assume that it is 
solved by some unspecified mechanism. } \cite{Nilles:1983ge}. This scenario is commonly called 
gravity mediation, and the CMSSM is one of its simplest and most popular 
realizations for phenomenological studies \cite{kkrw94}.
It is defined in terms of only five free parameters: 
common scalar ($\mzero$), gaugino ($\mhalf$) and 
tri--linear ($\azero$) mass parameters 
(all specified at the GUT scale) plus the ratio of Higgs vacuum 
expectation values $\tanb$ and $\text{sign}(\mu)$, where $\mu$ is 
the Higgs/higgsino mass parameter whose square is computed from 
the conditions of radiative electroweak symmetry breaking. In addition to these parameters, in this analysis we have the gravitino mass, $m_{3/2}$, that we keep as a free parameter.

The gravitino LSP (or NLSP) scenario in the CMSSM has been investigated 
thoroughly in the literature for the past years, both from the point of view
of cosmological implications for dark matter \cite{Roszkowski:2004jd} and 
for implications in collider searches \cite{Choi:2007rh}. 
One of the remarkable results of this scenario from cosmology is that the 
limits imposed on the reheating temperature of the Universe after 
an inflationary epoch have got down to a few $10^7$ GeVs \cite{Bailly:2009pe}. 
This constraint basically rules out the thermal leptogenesis mechanism as the 
mean to produce the observed baryon asymmetry of the universe. 

\begin{table}[t]
\centering
\begin{tabular}{|l | l l l | l|}
\hline
Observable & Mean value & \multicolumn{2}{c|}{Uncertainties} & ref. \\
 &   $\mu$ & ${\sigma}$ (exper.)  & $\tau$ (theor.) & \\\hline
$\delta a_\mu \times 10^{10}\; (e^+ e^-)$ &  29.5 & 8.8 & 2.0 & 
\cite{Miller:2007kk} \\
$\brbsgamma \times 10^{4}$ & 3.52 & 0.33 & 0.3 & \cite{hfag} \\
$\abundchi$ &  0.1099 & 0.0062 & $0.1\,\abundchi$& \cite{wmap5yr} 
\\\hline\hline
   &  Limit (95\%~\cl)  & \multicolumn{2}{r|}{$\tau$ (theor.)} & ref. 
\\ \hline
$\mhl$  & $>114.4\gev$  & \multicolumn{2}{r|}{$3 \gev$}
& \cite{Barate:2003sz} \\
Sparticle masses  &  \multicolumn{3}{c|}{As implemented in
Micromegas} & \cite{micromegas} \\ \hline
\end{tabular}
\caption{ \small Summary of the observables used in the analysis to 
constrain the CMSSM parameter space. Upper part:
Observables for which a positive measurement has been made. 
$\delta a_\mu= a_\mu^{\rm exp}-\amusm$ denotes the discrepancy between
the experimental value and the SM prediction of the anomalous magnetic
moment of the muon $\gmtwo$.
Lower part: Observables for which only limits currently
exist.
\label{tab:obs}}       
\end{table}

In this section, we apply the results of the previous sections for the 
degenerate gravitino scenario
to investigate the highest reheating temperature that can be reached in the 
CMSSM fulfilling the WMAP constraint on the CDM abundance and the relevant 
collider bounds; namely, direct SUSY searches, $(g-2)_\mu$ using 
($e^+e^- \rightarrow $hadrons) data\footnote{Notice that we obtain the hadronic contribution to  $(g-2)_\mu$ using only the data from ($e^+e^- \rightarrow $hadrons), inclusion of the $\tau$ data can decrease the discrepancy between the SM prediction and the experimental result.} and the $BR(B\rightarrow X_s \gamma)$. All 
the constraints imposed are summarized in Table~\ref{tab:obs}. 

We use the fortran package SUSPECT \cite{Djouadi:2002ze} to solve the RGEs
and to calculate the spectrum of physical sparticles and Higgs bosons, 
following the procedure outlined in \cite{Djouadi:2001yk}. 
To evaluate the CDM abundance in each point of the CMSSM parameter 
space, we employ the \texttt{MicrOMEGAs} code \cite{micromegas}.
The branching ratio for the $B \rightarrow X_s \gamma$ decay 
has been computed with the numerical code \texttt{SusyBSG} 
\cite{Degrassi:2007kj} using the full NLO QCD contributions, including 
the two-loop calculation of the gluino contributions presented 
in \cite{Degrassi:2006eh} and the results of \cite{D'Ambrosio:2002ex} 
for the remaining non-QCD $\tanb$-enhanced contributions. 
Finally we compute $\delta_{\rm had}^{\rm SM}a_\mu$ at full 
one-loop level adding the logarithmic piece of the quantum 
electrodynamics two-loop calculation plus two-loop contributions 
from both stop-Higgs and chargino-stop/sbottom \cite{Heinemeyer:2004yq}. 
Then, the effective two-loop effect due 
to a shift in the muon Yukawa coupling proportional to $\tan^2 \beta$ 
has been added as well \cite{Marchetti:2008hw}. Recall that the 
communication among the different codes is done via the SLHA 
accord \cite{Skands:2003cj}.

In particular, regarding the direct constraints on new particle searches, 
given that the theoretical error in computing the lightest 
Higgs mass $\mhl$ by SUSPECT is about $3 \, \gev$ \cite{Allanach:2004rh}, we require the calculated value of $\mhl$ to exceed $111 \, \gev$. In the case of 
observables for which a positive measurement has been made, we require our predictions to be within the $2 \sigma$ range, for which we have added the theoretical and 
experimental errors, found in Table~\ref{tab:obs}, in quadrature. 
In our numerical analysis we take $m_t = 173.1 \, \gev$ \cite{:2009ec}. 
For the case of the stau NLSP, in addition, we completely exclude points that do not satisfy the conservative 
bound on $Y_\stau < 10^{-14}$ \cite{Jedamzik:2007qk} from the 
catalyzed nuclear reactions \cite{CBBN} (see Fig.~\ref{fig:combinedstau}).


\begin{figure}[t]
\begin{center}
\includegraphics[angle=0,width=0.5\linewidth]{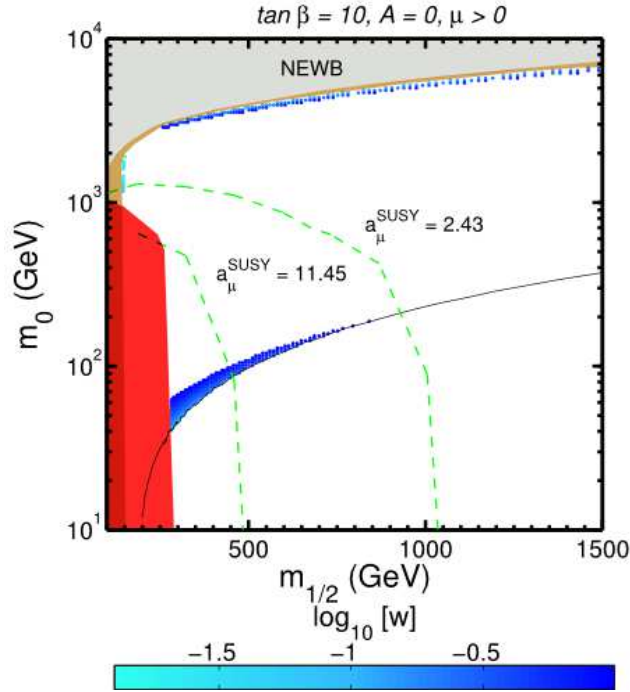}
\caption[text]{ \small Values of w in the $m_{1/2}$–-$m_0$ plane of the 
CMSSM parameter space for $m_t=173.1 \, \gev$, $A_0=0$, $\tanb=10$ and 
$\mu>0$. The red (medium grey) region is forbidden by the Higgs bound from LEP and in the very light grey region no correct electroweak symmetry breaking is obtained. The light brown (light grey) band below the NEWB area corresponds to the region forbidden by the LEP chargino bound. The regions below the dashed green lines satisfy the $(g-2)_\mu$ constraint at the 2 
$\sigma$ ($a_\mu^{\rm SUSY} > 2.43 \times 10^{-10}$) or 3 $\sigma$ level ($a_\mu^{\rm SUSY} > 11.45 \times 10^{-10}$).}
\label{fig:cmssm-tb10w}
\end{center}
\end{figure}

\begin{figure}[t]
\begin{center}
\includegraphics[angle=0,width=0.5\linewidth]{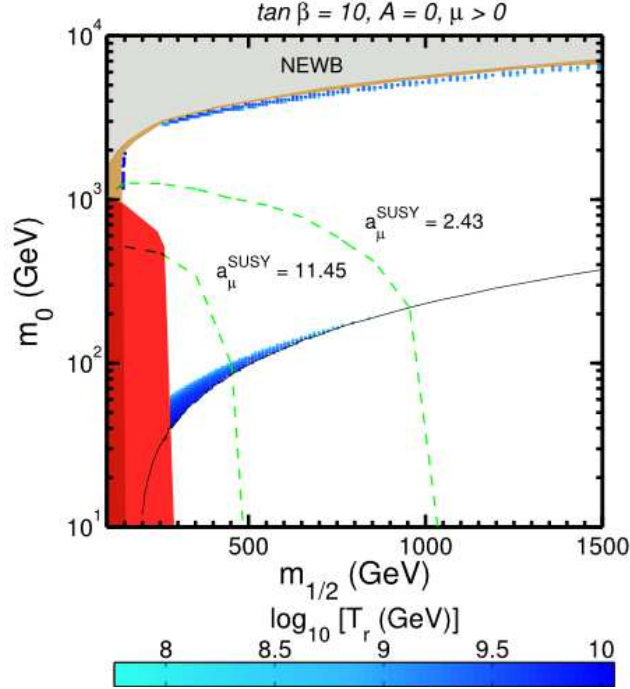} 
\caption[text]{ \small Maximal reheating temperatures in the $m_{1/2}$–-$m_0$ plane 
corresponding to the values of w in Fig.~\ref{fig:cmssm-tb10w}.}
\label{fig:cmssm-tb10tr}
\end{center}
\end{figure}

We are now ready to present some representative numerical results. We are going to calculate the abundance of the MLSP (MSSM LSP), ${\rm w} = Y_{\rm MLSP}/Y_{\rm CDM}$, at freeze-out. Notice that ${\rm w} = \omega$ if the MLSP is the NLSP, i.e. with gravitino LSP, but ${\rm w}= (1-\omega)$ when the gravitino is the NLSP. As a result, the reheating temperature will
be proportional to $(1-{\rm w})$ in both cases. 
In the following we will focus on three representative $\tanb$ values: one moderate, $\tanb=10$, and two large values, $\tanb=50$ and $\tanb=55$, and we 
take $A_0 =0$. Indeed, we have performed 
several other scans varying both $A_0$ and $\tanb$ and we found that there 
are no significant differences to these cases.

\subsection{Low-medium $\tan \beta = 10$}

In Fig.~\ref{fig:cmssm-tb10w} we show the region of ${\rm w} \leq 1$ (corresponding to $\Omega_{\chi}h^2 \leq 0.115$) in the 
($m_{1/2}$, $m_0$) plane for $A_0 = 0$, $\tanb = 10$, whereas we have chosen 
$\mu > 0$ motivated by $(g-2)_\mu$ data. 
In the CMSSM, the only mechanism which provides a dark matter abundance consistent with the WMAP value, if we require agreement with $(g-2)_\mu$ at 3 $\sigma$, is stau-neutralino 
coannihilations \cite{Nihei:2002sc}. 
This region is located  in a narrow band above the stau-neutralino 
degeneracy line ($\mstau= m_\chi$).
In addition to this relic abundance constraint, the only effective constraint 
in this region comes from direct Higgs searches 
at LEP which excludes low gaugino masses $m_{1/2} \lesssim 300 \gev$. In this figure, the allowed values for ${\rm w}$ are shown in different colours. 
As we can see, in this case, we can only obtain ${\rm w}$ between 1 and $\mathcal{O}(10^{-1})$. In the region of $\mstau\leq m_\chi$, although the annihilation 
mechanisms are more efficient than in the neutralino case due to fact that 
the stau is a charged particle, the lowest ${\rm w}$ we can reach is 
${\rm w}_{\min} \sim 7 \times 10^{-2}$, whereas from Eq.~(\ref{CBBNomega}) 
we see that the most conservative bound $Y_\stau < 10^{-14}$ from catalyzed 
nuclear reactions would require ${\rm w} \lesssim 2.7 \times 10^{-3}$. 
Hence, this region   of $\mstau\leq m_\chi$ is completely excluded for all values of $\Delta M$. If we eliminate the requirement of a non-zero SUSY 
contribution to   $(g-2)_\mu$, we obtain two small regions. First, there is  a vertical strip at high $m_0$ corresponding to a pole in the annihilation $\chi \chi \to b \bar{b}$ via the 
lightest Higgs. In this region we can obtain values of ${\rm w}$ as low as  ${\rm w} \simeq 0.01$. Then, there is a long strip, below the non-electroweak 
symmetry breaking (NEWB) region, at very large  $m_0$, where 
the neutralino is a mixed Bino-Higgsino state and the annihilation to $W^+ W^-$ is efficient enough to get a right relic abundance. However, in this region we obtain always ${\rm w} \gtrsim 0.1$.

Using these allowed values for w and the corresponding $\omega$ values, we obtain a bound on the mass difference $\Delta M$ in our degenerate gravitino scenario from Figure \ref{fig:combinedneutr}. According to Figure \ref{fig:combinedneutr}, for $\omega \geq 0.03$  the relevant constraint is diffuse gamma ray observations, and this requires $\Delta M \leq  10^{-2}$ GeV. This means that the lifetime of the NLSP (neutralino or gravitino) is longer than the age of the universe. 
Notice that, in the case of gravitino LSP, the fact that we cannot obtain w~$=\omega<0.1$ implies that this scenario would be completely ruled out by BBN, CMB and diffuse gamma ray observations, unless $\Delta M < 10^{-2}$ GeV and the neutralinos are beginning to decay at present. However, if the gravitino is the NLSP, we have two possibilities. The first possibility would be that the observed dark matter abundance is provided completely by the neutralino w~$\simeq 1$, with only a small fraction, at the level of $10^{-3}$, due to gravitinos. This corresponds to $\treh$ of the order of $10^7$ GeV, which is the usual situation in previous studies. The second possibility would be that there is a sizeable fraction of the dark matter due to gravitinos but, again, this would require $\Delta M < 10^{-2}$ GeV corresponding to a very long gravitino lifetime which permits to evade the cosmological bounds.   
 The maximum values of $\treh$ in these scenarios are shown in Fig.~\ref{fig:cmssm-tb10tr}. As it can be seen from this figure, values of $\treh > 10^9 \gev$ consistent with thermal leptogenesis are accessible in the model, although this is only possible if $\Delta M \leq 10^{-2}$ GeV. 

\subsection{Large $\tan \beta = 50$}
\label{tgb50}
\begin{figure}[t]
\begin{center}
\includegraphics[angle=0,width=0.5\linewidth]{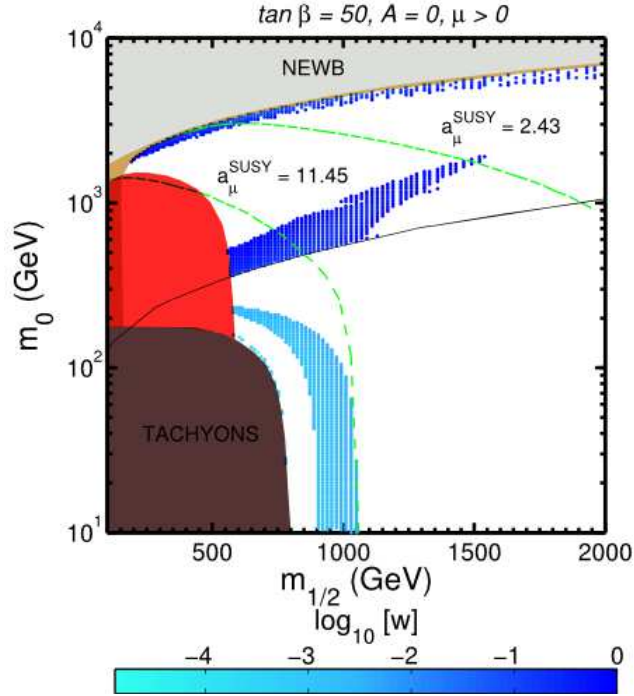} 
\caption[text]{ \small Values of w in the $m_{1/2}$–-$m_0$ plane of the 
CMSSM parameter space for $m_t=173.1 \, \gev$, $A_0=0$, $\tanb=50$ and 
$\mu>0$. In this case, the red region represents the bound from BR$(b \to s \gamma)$ at  $2\sigma$ and there is a new region in dark brown (dark grey) marked ``TACHYONS'' corresponding to the presence of tachyonic masses. The other colored regions have the same meaning as in the case of $\tan \beta =10$. Regions  below the dashed green lines satisfy the $(g-2)_\mu$ constraint at the 2 
$\sigma$ or 3 $\sigma$ level.}
\label{fig:cmssm-tb50w}
\end{center}
\end{figure}

In Fig.~\ref{fig:cmssm-tb50w}, we analyze the case of $\tanb = 50$. 
For large values of $\tanb$, in the neutralino MSSM LSP region ($m_\chi\leq \mstau$), we have different mechanisms to get a correct relic density consistent with WMAP. These mechanisms are i) stau-neutralino coannihilations, ii)  ``A-pole'' region,
where the s-channel exchange of the CP-odd Higgs boson, $A$, can become nearly 
resonant \cite{Ellis:2001msa, Roszkowski:2001sb} 
and iii) the ``focus point'' or ``hyperbolical branch'' where a significant 
higgsino component, enhances 
its annihilation cross sections into final states containing gauge 
and/or Higgs bosons \cite{Chan:1997bi, Feng:1999mn, Feng:1999zg}. 
In fact, the focus point region occurs at $m_0$ much larger than $m_{1/2}$ and  therefore multi-TeV scalar masses which implies that 
$\delta a_\mu^{MSSM}$ is well below the lower $2\sigma$ limit due to SUSY 
decoupling \cite{Moroi:1995yh}. However, if we accept $\delta a_\mu^{MSSM}$ at the $3\sigma$ level we can also reach the ``focus point'' region for $m_0 \simeq 2 \times 10^3 \gev$. 
Both the neutralino-stau coannihilation band and the ``A-pole'' region merge for $\tanb =50$ as we can see in this figure. This region is cut from the small $m_{1/2}$ values by the lower $2\sigma$ limit of the $\brbsgamma$ constraint \footnote{The reason for this is the destructive interference of the 
chargino/squark loops which grows with $\tanb$ \cite{Degrassi:2000qf}. 
This effect is enhanced for chargino masses $ \leq 100 \, \gev$ 
whereas is never dramatic for masses around $1$ TeV or larger.}. 
This constraint limits the minimum possible value of w obtainable through the ``A-pole''. Therefore we can only reach ${\rm w}_{\min} \sim 0.1$ both in the coannihilation and  ``A-pole'' regions. In the ``focus point'' region the situation is again similar and we can reach only values ${\rm w}_{\min} \sim 0.1$--$0.01$.

In the case of  $\mstau\leq m_\chi$, we can observe an allowed narrow band 
close to the tachyonic region in which the staus annihilate to  
$b \bar b$ via a pole in the s-channel exchanging a lightest Higgs. 
This process allows for values of ${\rm w}_{\min} \sim  10^{-5}$ which are 
consistent with the conservative CBBN constraint, 
$Y_\stau < 10^{-14}$--$10^{-15}$.  The broader band at large $m_{1/2}$ 
corresponds to the annihilation into a pair of light Higgses exchanging 
a light Higgs via a s-channel and here we 
obtain ${\rm w}_{\min} \sim  10^{-3}$ and therefore $Y_\stau < 10^{-14}$. 
On the other hand, we have to keep in mind that if we had imposed the more severe constraint $Y_\stau < 10^{-16}$, both regions would be ruled out.

Likewise, with these values for w, we obtain,  from Figure \ref{fig:combinedneutr}, the allowed values for the mass difference $\Delta M$. 
In the neutralino-stau coannihilation or A-pole regions only $\Delta M \leq 10^{-2}$ GeV are allowed, similarly to the $\tanb = 10$ case. However, in the region of  $\mstau\leq m_\chi$ with values of  w~$\gtrsim 10^{-5}$, we can see from Fig.~\ref{fig:combinedstau} that values of the mass difference, $90 \gev > \Delta M \gtrsim 1$ GeV,  would be still allowed for these points \footnote{We are considering only $\Delta M < 90 \gev$ where the hadronic BBN constraints are not efficient.}.

\begin{figure}[t]
\begin{center}
\includegraphics[angle=0,width=0.5\linewidth]{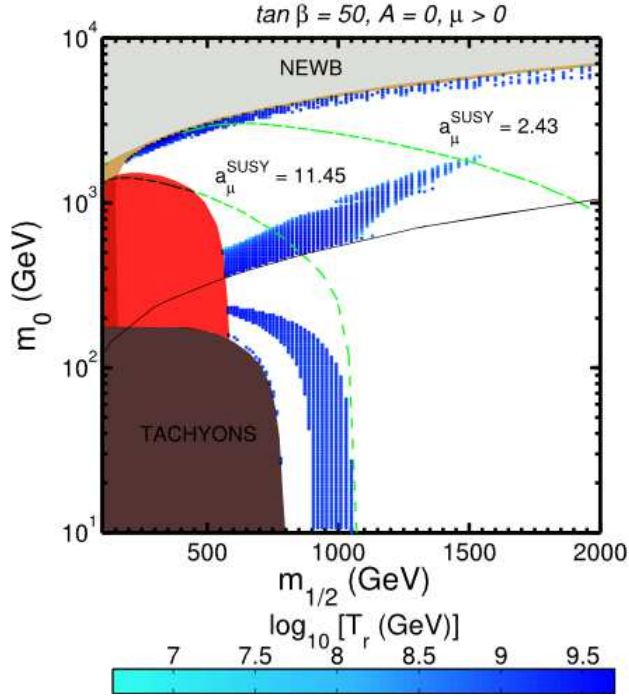} 
\caption[text]{\small  Maximal reheating temperatures in the $m_{1/2}$–-$m_0$ plane 
corresponding to the values of w in Fig.~\ref{fig:cmssm-tb50w}. Notice that the region below the stau--neutralino degeneracy line is not allowed in the case of gravitino NLSP.}
\label{fig:cmssm-tb50tr}
\end{center}
\end{figure}

The maximum values of $\treh$  are shown in Fig.~\ref{fig:cmssm-tb50tr}. As we have seen before, both for the gravitino LSP and gravitino NLSP cases $\treh \propto(1- {\rm w})$. The Maximal reheating temperatures for $m_{\chi}<m_\stau$ are always  $\treh > 10^9 \gev$ if the condition of $\Delta M \lesssim 10^{-2}$ GeV is satisfied, as in the case of $\tan \beta=10$. However, the region below the 
stau--neutralino degeneracy line are only allowed in the case of gravitino LSP when the maximal reheating temperatures are again $\treh > 10^9 \gev$, although 
in this case we do not require a tight degeneracy between gravitino and stau.

\subsection{Large $\tan \beta = 55$}

\begin{figure}[t]
\begin{center}
\includegraphics[angle=0,width=0.5\linewidth]{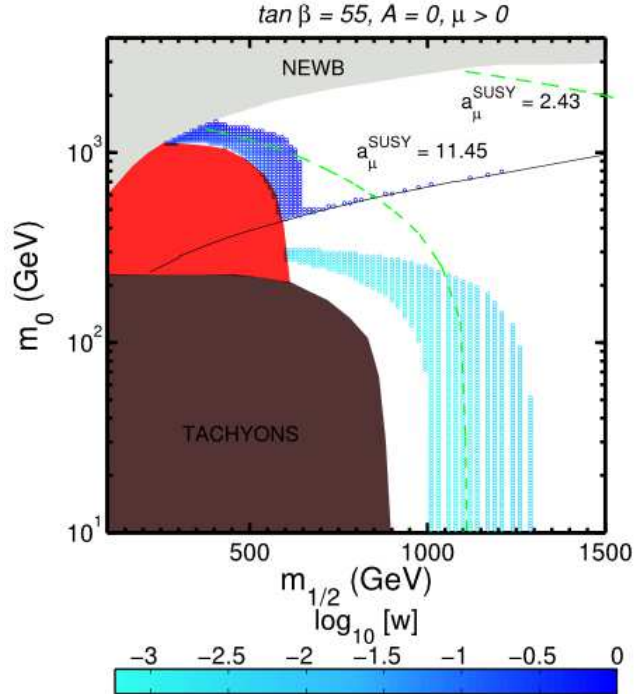} 
\caption[text]{\small Values of w in the $m_{1/2}$–-$m_0$ plane of the 
CMSSM parameter space for $m_t=173.1 \, \gev$, $A_0=0$, $\tanb=55$ and 
$\mu>0$. The meaning of the different regions is the same as in Fig.~\ref{fig:cmssm-tb50w}. The regions below the dashed green lines satisfy the $(g-2)_\mu$ constraint at the 2 
$\sigma$ or 3 $\sigma$ level.}
\label{fig:cmssm-tb55w}
\end{center}
\end{figure}

In Fig.~\ref{fig:cmssm-tb55w}, the case of $\tanb = 55$ is explored. 
Similarly to the case of $\tan \beta=50$, in the neutralino MSSM LSP region, the main annihilation mechanisms are
the neutralino-stau coannihilation and the ``A-pole'' resonance, however the ``focus point'' region is absent in this case.  Again the $\brbsgamma$ constraint limits the minimum possible value of w obtainable through the ``A-pole''. Therefore we can only reach ${\rm w}_{\min} \sim 0.1$, both in the coannihilation and ``A-pole'' regions.

In the case of  $\mstau\leq m_\chi$, as for $\tan \beta=50$, staus annihilate through an s-channel 
exchange of the light Higgs to a pair of light Higgses, although the annihilation to $b \bar b$ is suppressed in this case (it does not satisfy the required constraint $Y_\stau < 10^{-14}$). The allowed process, $\stau \stau \to h h $,  
allows for values of 
${\rm w}_{\min} \sim 6 \times 10^{-4}$ which are consistent with the conservative CBBN constraint for  $Y_\stau < 10^{-15}$. 

\begin{figure}[t]
\begin{center}
\includegraphics[angle=0,width=0.5\linewidth]{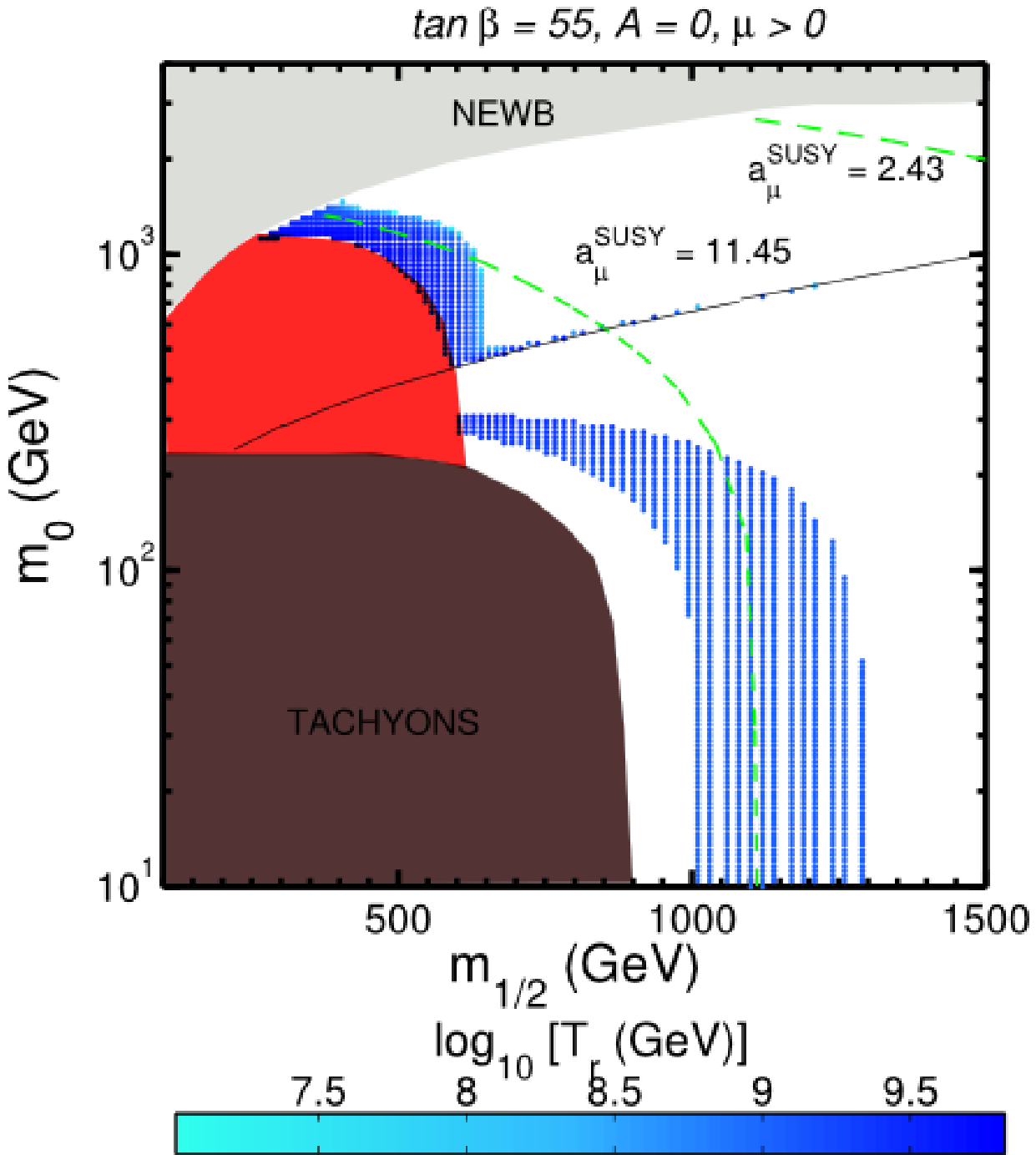}
\caption[text]{\small Maximal reheating temperatures in the $m_{1/2}$–-$m_0$ plane 
corresponding to the values of w in Fig.~\ref{fig:cmssm-tb55w}. As before the region below the stau--neutralino degeneracy line is not allowed in the case of gravitino NLSP.}
\label{fig:cmssm-tb55tr}
\end{center}
\end{figure}

Likewise, with these values of w, from Figure \ref{fig:combinedneutr} we obtain the allowed values for the mass difference $\Delta M$. 
As before, in the neutralino-stau coannihilation or A-pole regions only $\Delta M \leq 10^{-2}$ GeV are allowed and in the region of  $\mstau\leq m_\chi$  mass differences $90 \gev > \Delta M >2$ GeV would be still allowed. 
The corresponding values of $\treh$  are shown in Fig.~\ref{fig:cmssm-tb55tr}.
The discussion of section \ref{tgb50} applies also in this case.

\section{Conclusions}

\label{sec:conclusions}

Even though they are attractive candidates for cold dark matter,
gravitinos are usually a problem in standard cosmology. This is due to
the fact that they  typically decay at or just after the BBN putting
in danger the successful light-elements abundances. Requiring that
gravitinos  do not conflict with big-bang nucleosynthesis implies that
the reheating temperature should be low. In general, there is no
experimental constraint on how low $T_{\rm RH}$ should be, so in
principle a reheating temperature as low as MeV, so to permit BBN, is
perfectly allowed. However, successful thermal leptogenesis requires a
high reheating temperature $T_{\rm RH}\gtrsim 2 \times 10^9$ GeV. 

In this paper, we propose a solution that alleviates this tension by
making the gravitino degenerate with the lightest MSSM particle. This
has the direct consequence that the injected energy is suppressed,
making the decay products less dangerous for BBN. Due to the small
mass splitting, the gravitino and the MSSM lightest particle are
typically long-lived.  We analysed this scenario (the ``degenerate
gravitino'' scenario) by confronting it to cosmological and
astrophysical constraints. Since the NLSP decays at or after BBN, we
considered in addition to BBN, constraints from CMB spectral
distortions and diffuse gamma rays observations. First we performed a
model-independent analysis by considering a generic NLSP-LSP
degenerate scenario where the NLSP decays through NLSP$\to$LSP+ $X$
($X = \gamma, \tau$).  Since the final cold dark matter relic density
is the sum of both thermal and non-thermal contributions, we required
that the total cold dark matter  is consistent with cosmological
observation. Then, using the results of this analysis, we studied the
degenerate gravitino scenario in the context of the CMSSM where three
types of spectra arise, they are: gravitino NLSP with neutralino LSP
and gravitino LSP with neutralino or stau NLSP. Each of these cases
has been analysed in this framework defined by the usual high energy
parameters ($m_0$, $m_{1/2}$, $A_0$, $\tan\beta$, sign$(\mu)$) and the
gravitino mass $m_{3/2}$, and confronted with low-energy
observables. We find that high reheating temperatures consistent with
thermal leptogenesis can be found in all three scenarios if a sizable
part of cold dark matter comes from gravitinos produced at
reheating. In this case, depending on $\tan\beta$, we are led to
regions in the parameter space where the relic density of stau and
neutralinos are somewhat suppressed. In general to satisfy all the
constraints, the mass splitting between the NLSP and LSP should be
$\Delta M\simeq 10^{-2}$ GeV for the degenerate neutralino-gravitino
scenario, which implies very long-lived NLSPs which are beginning to
decay at present. On the other hand, in the  gravitino-stau scenario,
splittings in the range 10 GeV $\lesssim\Delta M \lesssim 90$ GeV are
still consistent with reheating temperatures of the order of $10^9$
GeV if we consider the conservative CBBN constraint $Y_{\rm CBBN} \leq
10^{-15}$.  

Let us comment on the required degeneracy in the ``degenerate gravitino'' scenario. Although a degeneracy of the order of $\Delta M\simeq 10^{-2}$ GeV certainly implies a certain amount of fine-tuning, this tuning is only two orders of magnitude stronger than the usual tuning required in the coannihilation or funnel regions to obtain the right relic density in the MSSM. On the other hand, notice also that the fine tuning in our scenario is much softer that the tuning required in other scenarios like inelastic dark matter \cite{TuckerSmith:2001hy}. 

Finally, it is also important to consider the phenomenological consequences of
this scenario in colliders. In the case of neutralino LSP or NLSP, the only
indirect signal of this scenario will be that the relic density of neutralinos inferred from the measurements of supersymmetric masses and couplings at LHC, will not match the observed cold dark matter abundance and will be smaller. 
However, the measurements at direct detection experiments will agree with the 
cross sections obtained from colliders.
On the other hand, for stau NLSP, the collider signatures would be spectacular, 
as the staus would be completely stable on collider scales and slow charged
tracks will appear in the detector \cite{champ}. Notice that similar signatures can arise in other scenarios like  
for instance in degenarate neutralino-stau scenario \cite{Kaneko:2008re} or in gauge-mediation scenarios. However, since the typical stau lifetimes 
in these scenarios are smaller than 1 sec, some of the staus will decay inside the detector. In contrast,
stau lifetimes range from $10^9-10^{15}$ seconds in our scenario,  making it very difficult to observe.  Nevertheless, following the analysis 
of \cite{Asai:2009ka}, stau lifetimes could be measured at LHC for mass splittings $30 \gev \lesssim \Delta M\lesssim 90 \gev$, corresponding 
to lifetimes  $10^{10} \sec \gtrsim \tau_\stau \gtrsim 10^8 \sec$. In this case, direct detection experiments will give a null results as all the dark matter at present times is made of gravitinos.  Therefore, the "degenerate gravitino" scenario will be probed at colliders and direct detection experiments if SUSY is discovered at LHC. \\[.0cm]

\noindent{\bf Note added:}~
While completing this work we noticed the preprint \cite{Peter:2010au}
where the astrophysical consequences in dark matter halo properties of
a neutral long lived decaying-dark matter particle were studied. The required
parameters in our ``degenerate gravitino'' scenario in the case of
neutralino LSP or NLSP are such that the constraints of  \cite{Peter:2010au} are
satisfied.

\section*{Acknowledgments} 

We thank T.~Moroi for useful discussions.
O.~V. and L.~B. acknowledge financial support from  spanish MEC and FEDER (EC) under grant FPA2008-02878 and Generalitat Valenciana under the grant PROMETEO/2008/004. 
O.~V was supported in part by European program MRTN-CT-2006-035482
``Flavianet''. L.~B. thanks the Abdus Salam ICTP, the Service de Physique Th\'eorique of ULB Brussels and the CERN-TH division for hospitality 
during the completion of this work.  
K.Y. Choi was partly supported by the Korea Research
Foundation Grant funded by the Korean Government (KRF-2008-341-C00008)
and by the second stage of Brain Korea 21 Project in 2006.
The work of R. Ruiz de Austri has been supported in part by MEC (Spain) 
under grant FPA2007-60323, by Generalitat Valenciana under grant 
PROMETEO/2008/069 and by the Spanish Consolider-Ingenio 2010 Programme 
CPAN (CSD2007-00042).
The use of the ciclope cluster of the IFT-UAM/CSIC is also acknowledged.

\end{document}